\documentclass[twocolumn,tighten,apjl,aas_macros]{aastex631}

\usepackage{amsmath}
\usepackage{natbib}
\usepackage{amsmath,amssymb}
\usepackage{xspace}
\usepackage{graphicx} % For figures
\usepackage{xcolor}
\usepackage{orcidlink}
\usepackage{hyperref}
\usepackage{amsthm}

\hypersetup{
    colorlinks=true,
    linkcolor=blue,
    citecolor=blue,
    anchorcolor=blue,
    bookmarks=true
}

\newtheorem{principle}{Principle} % For creating theorems

\newcommand{\modification}[1]{{\color{black} {#1}}}

\shorttitle{PTAs require hierarchical models}
\shortauthors{van Haasteren}

\begin{document}

\title{Pulsar Timing Arrays require hierarchical models}

\author{Rutger van Haasteren\orcidlink{0000-0002-6428-2620}}
\affil{Max-Planck-Institut f{\"u}r Gravitationsphysik (Albert-Einstein-Institut), Callinstra{\ss}e 38, D-30167, Hannover, Germany\\
Leibniz Universit{\"a}t Hannover, D-30167, Hannover, Germany}
\email{rutger@vhaasteren.com}

\begin{abstract}
     Pulsar timing array (PTA) projects have found evidence of a stochastic background of gravitational waves (GWB) using data from an ensemble of pulsars. In the literature, minimal assumptions are made about the signal and noise processes that affect data from these pulsars, such as pulsar spin noise. These assumptions are encoded as uninformative priors in Bayesian searches, though frequentist approaches make similar assumptions. Uninformative priors are not suitable for (noise) properties of pulsars in an ensemble, and they bias estimates of model parameters such as gravitational-wave signal parameters. Both frequentist and Bayesian searches are affected.
     In this article, more appropriate priors are proposed in the language of hierarchical Bayesian modeling, where the properties of the ensemble of pulsars are jointly described with the properties of the individual components of the ensemble. Results by PTA projects should be reevaluated using hierarchical models.
\end{abstract}


\section{Introduction}
\label{sec:intro}
Astronomy is on the cusp of a transformative era with the advent of gravitational-wave (GW) astronomy. The LIGO–Virgo collaboration has heralded this new age with observations in the higher-frequency bands \citep{ligoscientificcollaborationandvirgocollaborationObservationGravitationalWaves2016}, and now, pulsar timing arrays (PTAs) stand poised to complement these findings in the nanohertz regime with published evidence for a stochastic background of gravitational waves (GWB) \citep{agazieNANOGrav15Yr2023b,reardonSearchIsotropicGravitationalwave2023,antoniadisSecondDataRelease2023e, xuSearchingNanoHertzStochastic2023}.
This advancement holds the promise of reshaping our understanding of observational cosmology, galaxy formation, and physics of the early Universe.

PTAs represent a unique approach to GW detection. By precisely monitoring the arrival times of pulses from an array of millisecond pulsars \citep[MSPs,][]{backerMillisecondPulsar1982}, these projects seek to detect the subtle perturbations caused by passing GWs \citep{sazhinOpportunitiesDetectingUltralong1978,detweilerPulsarTimingMeasurements1979,romaniTimingMillisecondPulsar1989,fosterConstructingPulsarTiming1990}. The stability of MSP emissions is key to this method; however, this stability is not without some level of irregularity.
The rotational instabilities of the MSPs, often referred to as spin noise, presents a notable challenge for the PTA projects in their efforts to detect GWs \citep{cordesMeasurementModelPrecision2010,lentatiSpinNoiseSystematics2016}.

The complexity of analyzing an ensemble of pulsars simultaneously necessitates careful statistical modeling \citep{taylorNanohertzGravitationalWave2021}, particularly in the treatment of spin noise.
The PTA community has approached this with Bayesian inference, where separate spin-noise components with uninformative amplitude parameters for each MSP are added to the model. While this practice seems intuitive, it has sometimes led to posterior bias in GW analyses, where we use the term ``posterior bias'' to describe the apparent shift of the posterior distribution with respect to the true model parameter values. Posterior bias can be a consequence of model misspecification.
Multiple papers have pointed out that posterior bias occurs in contemporary analyses \citep{hazbounModelDependenceBayesian2020,zicEvaluatingPrevalenceSpurious2022,johnsonGravitationalwaveStatisticsPulsar2022}, but no exact cause or satisfactory solution was given. These warning signals have been mostly ignored by subsequent searches for GWs. Those same priors are still used today, such as in recent high-profile publications \citep{agazieNANOGrav15Yr2023b,reardonSearchIsotropicGravitationalwave2023,antoniadisSecondDataRelease2023e, xuSearchingNanoHertzStochastic2023}.

Even though currently used uninformative priors have been confirmed to work with injection recoveries in the literature \citep[e.g.][and most other methods papers on GWB detection]{lentatiHyperefficientModelindependentBayesian2013,vanhaasterenMeasuringGravitationalwaveBackground2009,ellisEfficientApproximationLikelihood2013,johnsonNANOGrav15yearGravitationalWave2023a}, no rigorous statistical assessment of (posterior) bias has been possible. What has been done are studies that focus on the validity of the posterior distribution and its implementation, which can be done with $p$-$p$ plots of ensembles of mock data \citep[e.g.][]{ellisEfficientApproximationLikelihood2013}.
However, for those $p$-$p$ plots to be consistent, one needs to use the \emph{same} priors for data generation as for analysis. By doing that, one glosses over a significant source of posterior bias: the use of uninformative priors on an \emph{ensemble} of observable quantities that follow a different distribution in reality than assumed in the prior \citep{priestleyEvolutionarySpectraNonStationary1965}. \citet{goncharovConsistencyParkesPulsar2022}
pointed out that this type of model misspecification is the cause for posterior bias, and in a way the current paper generalizes their proposed solution.

Frequentist estimators are similarly affected. A minimum variance unbiased estimator \citep[MVUE;][]{anholmOptimalStrategiesGravitational2009,chamberlinTimedomainImplementationOptimal2015,allenHellingsDownsCorrelation2023a} needs to assume the data distribution is known, which requires knowledge of the noise parameters.
Or the posterior distribution is used to marginalize over the noise parameters while evaluating the optimal statistic \citep{vigelandNoisemarginalizedOptimalStatistic2018, sardesaiGeneralizedOptimalStatistic2023a,vallisneriPosteriorPredictiveChecking2023}, which in turn requires the same uninformative priors.
Fortunately, the MVUE for an isotropic stochastic background of GWs that does not use pulsar autocorrelations is still unbiased by construction, though no longer minimum variance if suboptimal assumptions are used.

In this letter, we propose a shift in perspective, advocating for the adoption of hierarchical Bayesian modeling (HBM) in PTA analysis. This approach has multiple benefits over currently used priors, chief among them being that the GW parameters will exhibit less posterior bias.

In Section \ref{sec:uninformative_priors} we present a toy model that demonstrates the posterior bias of currently used uninformative priors. In Section \ref{sec:hierarchical_modeling} we discuss the problems of uninformative priors theoretically, and introduce Hierarchical Bayesian Modeling that should replace them. In Section \ref{sec:simulations} we give some more realistic examples, after which we briefly discuss current attempts in the literature to address posterior bias in Section~\ref{sec:litcomparison} I. The implications of this letter on the PTA community are discussed in Section~\ref{sec:implications}, after which we end with concluding remarks and an outlook in Section~\ref{sec:conclusions}.

\section{Current prior use in PTA projects}
\label{sec:uninformative_priors}
When observing and modeling the lighthouse-like pulses of electromagnetic radiation that pulsars emit, a detailed \emph{timing model} \citep{edwardsTempo2NewPulsar2006,luoPINTModernSoftware2021} with multiple components and parameters is fit to the data.
Any realistic analysis needs to account for the systematic effects that the timing model introduces. This section introduces a toy model that is unhindered by the complications of reality, as the effects of the timing model have no bearing on the way the prior influences inference on GW and other model parameters.

\begin{table*}[ht!]
    \centering
    \begin{tabular}{|c|c|c|c|c|}
        \hline
         & \textbf{Box} & \textbf{Normal} & \textbf{Identical} & \textbf{BiModal} \\
        \hline
        Detector noise $\log_{10}(\sigma_a)$ & $\sim \text{Uniform}(1.2,2.2)$ & $\sim \mathcal{N}(2.5,2^2)$ & $=1.4$ & $\sim \mathcal{N}(0.5,1) + \mathcal{N}(4.5,1)$ \\
        \hline
        Signal power $\log_{10}(h)$ & $=1.0$ & $=1.0$ & $=1.0$ & $=1.0$ \\
        \hline
    \end{tabular}
    \caption{Four scenarios for the underlying distributions of model parameters of the Toy Model. These values were chosen such that the signal was somewhat comparably detectable in all datasets.}
    \label{tab:scenarios}
\end{table*}

\subsection{Toy problem}
\label{sec:toy_model}
The toy model of this section is a simplification of actual PTA data. Assume we have data from $N_p$ detectors each associated a random position on the sky, and each detector yields $N_o$ observations. There are only two processes that give a response in these detectors: the noise process $X_{n}$ and the signal process $X_{s}$. The process $X_{n}$ produces IID (independent identically distributed) data in each detector, Gaussian distributed. The amplitude of $X_{n}$ is specific to the detector:
\begin{equation}
    n_{ai} \sim \mathcal{N}(0, \sigma_{a}^{2}),
    \label{eq:noise}
\end{equation}
where $n_{ai}$ is the $i$th noise process realization at detector $a$, where we use the convention that $i$ labels observations $i \in [1,N_o]$ and $a$ labels detectors $a \in [1,N_p]$. The standard deviation of the IID data in detector $a$ is given by $\sigma_{a}$. we denote a Gaussian distribution with mean $\mu$ and standard deviation $\sigma$ as $\mathcal{N}(\mu,\sigma^2)$.

The signal process $X_{s}$ is similar to $X_{n}$ in that it is a random process that, when considering only a single detector, produces IID data according to $\mathcal{N}(0, h^2)$. However, when considering an array of detectors, the realizations of $X_{s}$ are correlated.
The correlations that PTA projects focus on the most are so-called Hellings \& Downs correlations, which represent the average correlations induced by an isotropically unpolarized ensemble of GW sources across the sky \citep{hellingsUpperLimitsIsotropic1983}. These correlations $\Gamma(\zeta_{ab})$ only depend on the angular separation $\zeta_{ab}$ of detector $a$ and detector $b$:
\begin{eqnarray}
  \Gamma \left(\zeta_{ab}\right) &=& \Gamma_{\rm{u}}\left(\zeta_{ab}\right) + \delta_{ab}\Gamma_{\rm{u}}(0) \nonumber \\
  \xi_{ab} &=& \frac{\left(1 - \cos\zeta_{ab}\right)}{2} \nonumber  \\
  \Gamma_{\rm{u}}\left(\zeta_{ab}\right) &=& \frac{3\xi_{ab} \log\xi_{ab}}{2} - \frac{1}{4}\xi_{ab} + \frac{1}{2}, 
\end{eqnarray}
where $\delta_{ab}$ is the Kronecker delta. The presence of the $\delta_{ab}$ term represents the so-called \emph{pulsar term}, which describes effects of the GWs on the detector that are uncorrelated between pulsars.
The notation $\Gamma_{\rm{u}}(\zeta_{ab})$ comes from \citet{allenVarianceHellingsDownsCorrelation2023a}, where \emph{average} correlations $\Gamma$ are derived from an \emph{unpolarized} isotropically distributed background of GWs. With this, the contributions of $X_{s}$ to the detectors become:
\begin{eqnarray}
    s_{i} &\sim& \mathcal{N}(0, \mathbf{\Sigma}) \nonumber \\
    \Sigma_{ab} &=& h^2 \Gamma(\zeta_{ab})
\end{eqnarray}
where $h$ denotes the signal amplitude, similar to the meaning of $\sigma_{a}$ for the noise, and $\mathbf{\Sigma}$ is the covariance of the multivariate Gaussian distributed variable $s_{i}$. The elements of $s$ of detector $a$ and observations $i$ are written as $s_{ai}$. The full data $d$ is then
\begin{equation}
    d_{ai} = s_{ai} + n_{ai}.
    \label{eq:data}
\end{equation}

The only free model parameters in the toy model are $\sigma_{a}$ and $h$. The experimental setup is defined by the number of detectors $N_p$ and the number of observations per detector $N_o$. Given the numerical values of those parameters, it is possible to generate mock \emph{realizations} of $d$.
In a realization of our mock universe, the values of $\sigma_a$ follow the distribution that represents the accuracy of a detector in our ensemble of detectors. For PTAs, this distribution represents the variation in pulsar spin stability in our Galactic population of observable millisecond pulsars. We consider four scenarios, listed in table \ref{tab:scenarios}, where the distribution of detector noise comes from specific distributions. These signal is of the same amplitude \emph{in expectation} for all scenarios, and the expected noise amplitudes of Table~\ref{tab:scenarios} are tweaked so that the signal is detectable. In what follows, these different distributions will serve as examples to test the priors on the detector noise.

\subsection{Analysis using uninformative priors}
\label{sec:toy_model_results}

\begin{figure}
    \includegraphics[width=0.48\textwidth]{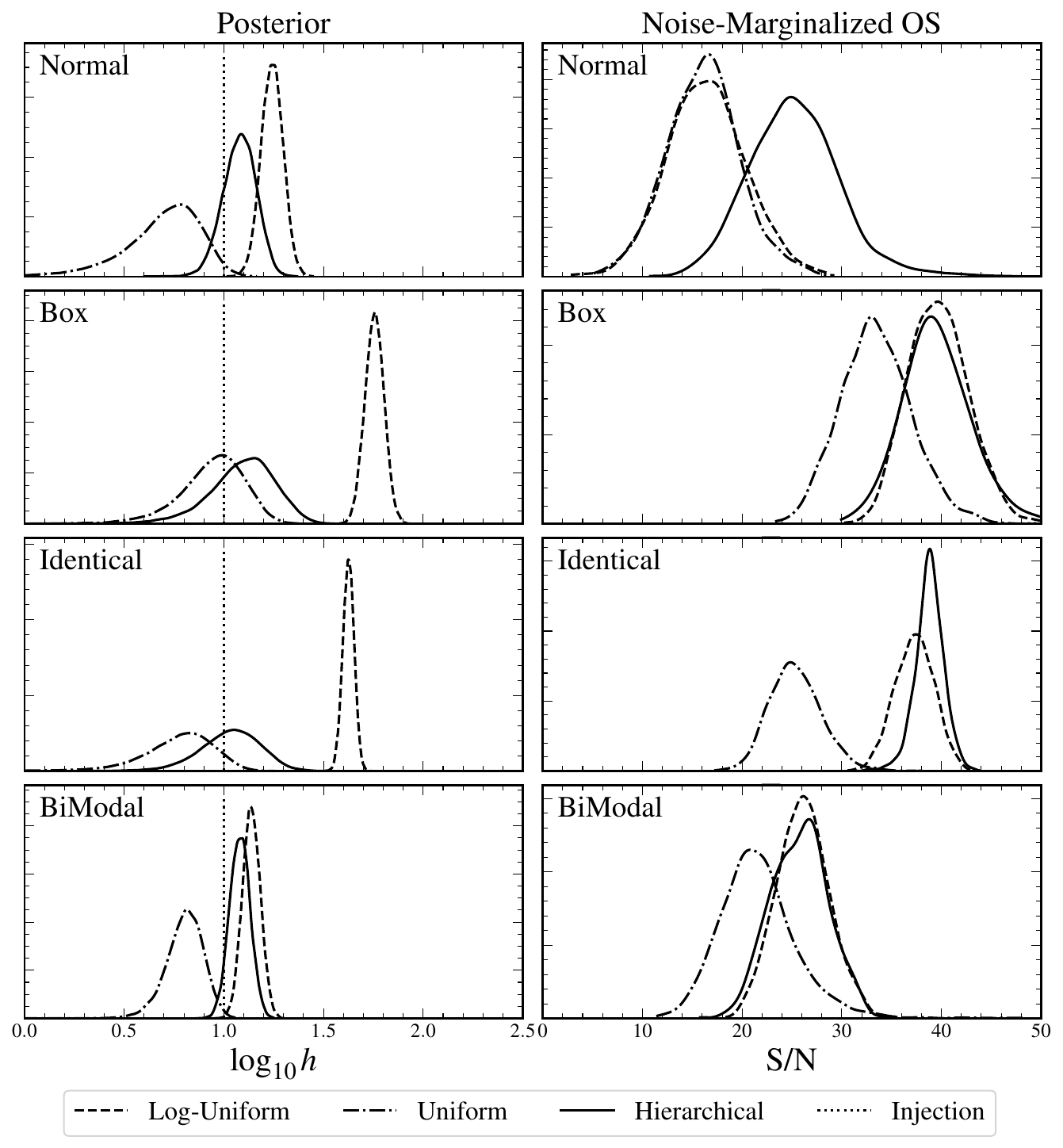}
    \caption{Posteriors and Optimal Statistic S/N for a realization for each of our four scenarios. Many realizations were made and analyzed (see Figure~\ref{fig:ppplots}), the graphs shown here are for the first realization of each scenario.
    Left columns are the $P(\log_{10}h^2 \mid d)$ posterior distributions for the model with Hierarchical, Log-Uniform, and Uniform priors on the detector noise amplitudes. The right columns are the normalized distributions that one gets for the values of the Optimal Detection statistic S/N, evaluated using noise parameters from random samples drawn from the posterior distribution. The rows are for a specific realization of the Normal, Box, and Identical scenarios. The signal prior is always the log-uniform prior given by Equation~\eqref{eq:luh}}
    \label{fig:scenario-posteriors}
\end{figure}

The toy model introduced in this section captures a couple of important characteristics of current PTA projects: 1. the signal of interest $s$ is a correlated random process with unknown amplitude, 2. every detector has detector noise with unknown amplitude that varies from detector to detector, 3. the signal and the detector noise look identical aside from the correlations between detectors, and 4. many detectors are necessary to distinguish signal from noise. 
It is only a few lines of code to produce a model that analyzes this kind of data in probabilistic programming languages like \texttt{PyMC} \citep{salvatierProbabilisticProgrammingPython2015}. The currently-used uninformative priors that are being used in the PTA community for detecting a GWB are log-uniform priors, which are defined as:
\begin{eqnarray}
    \log_{10}\sigma_a^2 &\sim& \rm{Uniform}(-5, 5) \nonumber \\
    \log_{10}h^2 &\sim& \rm{Uniform}(-5, 5)
    \label{eq:luh}
\end{eqnarray}
Similarly, the often-used uniform priors which are commonly used to place an upper-limit on variables are defined as:
\begin{eqnarray}
    \sigma_a^2 &\sim& \rm{Uniform}(0, 10^{5}) \nonumber \\
    h^2 &\sim& \rm{Uniform}(0, 10^{5})
\end{eqnarray}
In this section, both types of uninformative priors are used to analyze a single realization for each of the scenarios of Table \ref{tab:scenarios}. The data are generated using \texttt{numpy} \citep{harrisArrayProgrammingNumPy2020} routines, and they are analyzed with \texttt{PyMC}.
The (marginalized) posterior distributions for both Log-Uniform and Uniform priors are shown in Figure \ref{fig:scenario-posteriors}. Signal priors on $\log_{10} h$ have been the log-uniform priors given by Equation~\eqref{eq:luh} for all plots. Switching to Uniform priors on the signal amplitude $h^2$ would not change the results significantly.
In the case of Log-Uniform priors for the detector noise, the estimate of $h$ is systematically high, whereas the estimate of $h$ is generally low when using Uniform priors. All this is pretty intuitive: favoring larger detector noise decreases the recovered signal amplitude $h$, because more of the variance in the data is attributed to noise.

These results are somewhat sobering, and given the analogous use of priors in PTA projects this exposes a shortcoming of current analyses of the PTA community. Although several papers \citep{goncharovConsistencyParkesPulsar2022,zicEvaluatingPrevalenceSpurious2022,johnsonGravitationalwaveStatisticsPulsar2022} have raised the alarm regarding the possibility of posterior bias in current PTA analyses, these issues have not been addressed properly. In fact, the most recent high-profile publications in the PTA community have still used the same assumptions on the ensemble of pulsars by using uninformative priors.
Moreover, other studies that have been carried out to test the consistency of PTA analysis methodologies in the literature have only tested the validity of the posterior distribution. As such, these studies have heavily relied on visualizations of $p$-$p$. Those type of analyses only work if the posterior distribution uses the same priors and the injections. And, of course, if the analysis prior and the injection prior are equal, the prior choice challenge is avoided.

\section{Hierarchical modeling}
\label{sec:hierarchical_modeling}

PTA projects use data from an ensemble of pulsars, where signal and noise processes of TOAs are modeled meticulously with great precision. These pulsars come from the collection of observable Galactic MSPs, which means that the noise parameters of the MSPs come from some underlying distribution. By analyzing the data of the ensemble of pulsars jointly, it becomes possible to model this underlying distribution of model parameters. Current analyses by PTAs purposely refrain from modeling this distribution, causing a mismatch between the generative model for the data and the model used for the posterior distribution. The results in Section~\ref{sec:toy_model_results} show that this causes problems with inference on parameters that are covariant with the ensemble of noise parameters as a whole (i.e. the GW signal $h$).
This covariance makes spin noise of particular interest. Even though for instance dispersion measure variations and pulse profile variations are also dominant sources of noise for some pulsars \citep[e.g.][]{agazieNANOGrav15Yr2023a}, it is the ensemble of spin noise parameters that is most covariant with a potential GW signal, and it is therefore the mismatch of the prior and the generative model of the spin noise that lies at the heart of the posterior biases that have been found in the literature.

In this section we propose Hierarchical Bayesian Modeling (HBM) as a solution to the situation PTAs are in, where apparent posterior bias results from using uninformative priors on the noise parameters. In some other areas of science HBMs have been adopted widely, and there is a large body of literature to draw from \citep[for a first example in GW science, look up][]{adamsAstrophysicalModelSelection2012}.
For a more general and in-depth discussion regarding multilevel and hierarchical Bayesian models, we refer to Section 3 of \citet{loredoMultilevelHierarchicalBayesian2019}, who do an excellent job introducing all the subtleties of HBMs in astrophysics.

\subsection{Modeling the underlying distribution}
\label{sec:underlyingdist}

When an ensemble of pulsars is analyzed, the data are providing us with insight in the underlying distribution that describe the parameters of the ensemble. If we are interested in spin noise, there are enough pulsars in our array to inform us on the distribution of spin noise in our Galaxy. If we are interested in pulsar distance and position in the Galaxy, then the Astrometry parameters of the ensemble of pulsars are informative on those distributions. This type of modeling of an underlying distribution as an extra ``level'' in our analysis is often referred to as multi-level modeling. The Bayesian version of this is referred to as a Hierarchical Bayesian Model (HBM).

There are many examples of these kind of problems across science \citep{gregoryBayesianLogicalData2005,hoggDataAnalysisRecipes2010,siviaDataAnalysisBayesian2006,hoggINFERRINGECCENTRICITYDISTRIBUTION2010,loredoMultilevelHierarchicalBayesian2019}. In some of these, biases are exposed and addressed, while in others variance of the estimators is reduced.
Interestingly, the point estimators for the parameters of single items out of an ensemble based on a multi-level model are often biased estimators. This makes sense: the estimator value is typically ``pulled'' towards the center of the underlying distribution. Said differently, the pooling of information allows the individual estimates to borrow strength from one another. Although biased, the estimators from a multi-level model have attractive properties, such as reduced variance. In the case of PTAs, posterior bias in covariant model parameters (the GWB parameters) is greatly reduced. Said more generally:

\begin{principle}
\modification{A model that properly represents a physical system in its entirety, needs to describe the distribution of parameters of an ensemble if that ensemble is part of the physical system.}
\end{principle}

\modification{This is a mathematical statement rooted in Bayes Theorem. By not describing the underlying distribution, one would have a misspecified model, and parameter inferences can become biased. Any model parameters that are covariant with the ensemble parameters will, in turn, also become subject to posterior bias.}

\subsection{The Log-Normal HyperPrior}
\label{sec:lognormalprior}

By modeling the underlying distribution in our joint analysis of detector data, an HBM is constructed that correctly captures both the statistics of the individual detector noise and the distribution that describes the variations from detector to detector. A good first guess for the prior on the noise amplitude is a log-normal distribution.
This causes both the scale and the spread of the noise amplitude to be inferred with minimal assumptions. This changes the prior for the noise amplitude:

\begin{eqnarray}
    \label{eq:hierarchical_model}
    \log_{10}\mu_{\sigma} &\sim& \mathcal{N}(1, 6^2) \nonumber \\
    \log_{10}\sigma_{\sigma} &\sim& \rm{Uniform}(0.05, 4) \\
    \log_{10}\sigma_{a} &\sim& \mathcal{N}\left(\log_{10}\mu_{\sigma}, \left(\log_{10}\sigma_{\sigma}\right)^2\right) \nonumber \\
    \log_{10}h &\sim& \rm{Uniform}\left(-5, 5\right) \nonumber .
\end{eqnarray}
Note that we are free to choose the priors on the hyperparameters $\mu_\sigma$ and $\sigma_\sigma$. These particular choices are convenient, practical, and sufficient for the present discussion.
Note that $\log_{10}\sigma_\sigma$ is the standard deviation of the prior distribution on the ensemble of noise parameters, not the prior on the individual noise parameters, denoted by $\log_{10}\sigma_{a}$. Together with Equation~\eqref{eq:noise}--\eqref{eq:data}, Equation~\eqref{eq:hierarchical_model} gives a full specification of the model. Analyzing the same dataset as above with this model, we obtain the results in Figure \ref{fig:scenario-posteriors} labeled as ``Hierarchical''. The resulting posteriors with the Hierarchical prior are universally more consistent with injections.

Aside from checking for posterior bias by eye in Figure \ref{fig:scenario-posteriors}, we make use of two methods to select the best model for my datasets. The first one is Bayesian model selection. Bayes Factors can be used to select between sets of assumptions and models about the data, which means that different prior assumptions can be compared. The Bayes factors of HBM vs Log-Uniform for the ``Normal'', ``Box'', and ``Identical'' scenarios from Figure~\ref{fig:scenario-posteriors} are respectively $\log\mathcal{B} = 33.6$, $\log\mathcal{B} = 33.5$, and $\log\mathcal{B} = 19.6$.

The second method to pick the best model for the datasets in this section is based on the fact that this is a toy model with mock data, so we can generate many realizations of data for which the injected parameters are known. A $p$-$p$ plot can then be used to check for consistency, which is done in Section \ref{sec:ppplots}.

\subsubsection{Qualitative discussion}
\label{sec:qualitativediscussion}

In the Pulsar Timing community, the log-uniform prior has been argued for from the perspective of the likelihood: where is the likelihood high in parameter space, and what happens if we sample the posterior? Therefore, it is instructive to reflect on what happens with hierarchical priors compared to static uninformative priors. The main difficulty, conceptually, is that the parameter space becomes high-dimensional when an ensemble of parameters is involved. It is not easy to visualize volumes and densities in more than three dimensions. For instance, in a hyper-sphere of $n$-dimensions, all volume is concentrated at the outer shell when $n\rightarrow\infty$, and in a similar hyper-cube the volume is concentrated at the corners.

A useful way to think about Bayesian models comes from the approach of Nested Sampling \citep{skillingNestedSampling2004} where the fully marginalized likelihood integral (FML, also called the \emph{evidence}) is transformed from an integration over all model parameters to an integral over the one-dimensional quantity of the \emph{prior volume}, like one would with Lebesgue integration \citep{evansMeasureTheoryFine1991}. When one is carrying out ordinary MCMC simulations, the \emph{typical set} is the region of parameter space where the MCMC sampler spends most of its time sampling from. An important consideration here is how large of a fraction of the total prior volume is taken up by the typical set. The better the model, the larger this fraction is.

In Nested Sampling the parameter space is usually transformed to the unit hypercube for simplicity, where all parameters run in the range $x \in [0,1]$. The transformation is done for all parameters using the percentage point function (PPF). On the hypercube, the FML is simply the average likelihood. Another way to say this is that the FML is the likelihood averaged over the prior distribution. The FML is a measure for how well the model fits the data. A model that fits the data well does not have a high maximum likelihood, but a high average likelihood over the entire prior distribution. This means that for such a model, high likelihood values take up a large fraction of prior volume compared to competing models. Conversely, poorly chosen priors can cause high likelihood values to occur in such a tiny fraction of the total prior volume that the typical set shifts elsewhere: there is so much prior volume elsewhere that the high likelihood region does not contribute significantly to the FML.

To simplify, the difficulty with uninformative priors is that, by construction, high likelihood values only occur in a fraction $f<1$ of their prior volume. If an ensemble of $n$ uninformative priors is collectively chosen, the prior volume where high likelihood values occur is reduced to $f^n$. This is not necessarily cause for concern if the parameters governed by that prior are not covariant with other parameters of interest (e.g. PTA white noise parameters). However, if the parameters are covariant with, say, the amplitude of the stochastic GW background, the choice of prior greatly influences the prior volume. Note also that the fraction $f$ typically depends arbitrarily on the boundaries that were set on the prior; a typical scenario for otherwise improper priors.

\begin{figure}
    \includegraphics[width=0.48\textwidth]{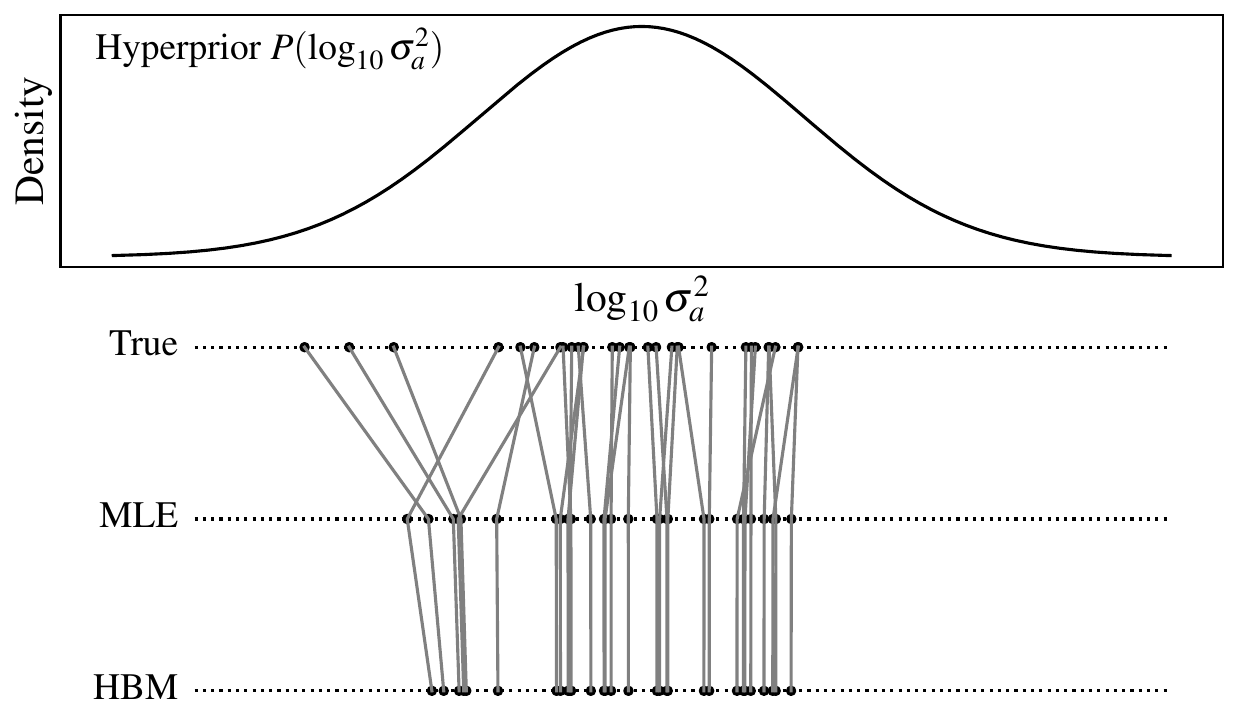}
    \caption{Example of shrinkage for the ``Normal'' scenario of Figure~\ref{fig:scenario-posteriors}. The top panel displays the hyperprior --- the underlying distribution of the ensemble of detector noise $\log_{10}\sigma^2_a$. Underneath, on the ``True'', the points represent the actual noise levels of $30$ detectors in the ensemble. On the ``MLE'' line, the Maximum-Likelihood Estimators of those detector noise levels are shown. On the ``HBM'' line, the noise estimates of the Hierarchical Bayesian Model are shown. Shrinkage is visible on the outer points, which are regressed slightly to the center of the hyperprior. The Mean-Squared Error of the HBM are lower than for the MLE.}
    \label{fig:shrinkage}
\end{figure}

A hierarchical prior, like the log-normal prior introduced in this letter, introduces the flexibility for the parameter priors to be fit to the data through the hyperparameters that shape those priors. Conditioned on the hyperparameters, this then means that the high likelihood values occur in a much larger fraction of the prior volume. Said differently:

\begin{principle}
    \modification{Under an HBM, the typical set of the low level parameters becomes a large fraction of the corresponding prior volume.}
\end{principle}
    This does several things:
1. this increases the FML, meaning that a hierarchical Bayesian model will have a Bayes factor $>1$ over models with uninformative priors.
2. it decreases potential posterior bias stemming from too small a prior volume being occupied by high likelihood values,
3. the parameter priors of the ensemble are no longer improper,
4. we gain information about the underlying distribution, and
5. the parameter estimates for items of the ensemble borrow strength from one another through pooling of information.
That last point is subtle: in our toy model this means that the ensemble of detectors jointly are informative regarding the noise in a single detector. 
This phenomenon is often referred to as \emph{shrinkage} \citep{gelmanBayesianDataAnalysis2013, loredoMultilevelHierarchicalBayesian2019}. The ``Normal'' scenario of Figure~\ref{fig:scenario-posteriors} also exhibits shrinkage: see Figure~\ref{fig:shrinkage}. There are many other subtleties with HBMs that are out of the scope of this letter. Please refer the citations above for a much more in-depth discussion.

\subsubsection{Comparisons using p-p plots}
\label{sec:ppplots}
Armed with an ensemble of realizations of mock data, it is possible to carry out a systematic assessment of consistency between data and model. Let's start with Bayes Theorem:
\begin{equation}
    P(\theta \mid d, H) = \frac{1}{Z} P(d \mid \theta, H) P(\theta \mid H),
    \label{eq:bayes}
\end{equation}
where $Z=P(d \mid H)$ is the FML/evidence,  $P(\theta \mid d, H)$ is the posterior distribution under model hypothesis $H$, $P(d \mid \theta, H)$ is the likelihood, and $P(\theta \mid H)$ is the prior distribution for the model parameter $\theta$. We only focus on a single model parameter here, so these distributions are one-dimensional. The cumulative posterior distribution function 
\begin{equation}
    C(\theta \mid d, H) = \int_{-\infty}^{\theta} \rm{d}\theta^\prime P(\theta^\prime \mid d, h),
\end{equation}
maps our parameter $\theta$ to the unit interval $[0,1]$. The procedure of generating realizations $d_r$ of data is as follows
\begin{itemize}
    \item $\theta_{\rm{inj},r} \sim P(\theta \mid H)$
    \item $d_r \sim P(d \mid \theta_{\rm{inj},r}, H)$
\end{itemize}
where $r \in [1, N_r]$ labels the realization of data $d_r$. First the injection parameters $\theta_{\rm{inj},r}$ are drawn from the prior, after which the data are generated by drawing it as a random variable from the likelihood conditioned on the model parameters. Note how both steps taken together represent the right-hand side of Bayes Theorem in Equation~\eqref{eq:bayes}. Now define $\xi_{r}$ as:
\begin{eqnarray}
    \xi_{r} &:=& C(\theta_{\rm{inj},r} \mid d_r, H) \\
    \xi_{r} &\sim& \rm{Uniform}(0,1)
\end{eqnarray}
where the last line follows from the first one, since it represents the posterior of the left-hand side of Bayes Theorem of Equation~\eqref{eq:bayes}. It is important to note that this is only true under hypothesis $H$. Note how $Z=P(d \mid H)$ on the right-hand side of Bayes Theorem indeed refers to the data conditioned on $H$.
Therefore, if the data were not generated through the generating process defined by $H$ above, then $\xi_{r}$ would not be uniformly distributed. This can be tested statistically in multiple ways. A common one is to order the $\xi_{r}$ as $\xi_{r^\prime}$ such that $\xi_1 \le \xi_2 \le \ldots \le \xi_{N_r}$, and to define $p_{r^\prime} = r^\prime/N_r$. Taken individually, the $\xi_{r^\prime}$ are now binomially distributed: $\xi_{r^\prime} \sim \rm{Binomial}(p_{r^\prime}, N_r)$.
Therefore, by checking whether $\xi_{r^\prime}$ is consistent with draws from the binomial distribution, it is possible to assess whether our posterior distribution is consistent with the injection parameters. Note that the $\xi_{r^\prime}$ are not independent draws from the binomial distribution, so it is not justified to do $N_r$ tests on the Binomial distribution. Examples of tests that can properly take the dependence into account are the Kolmogorov-Smirnov test \citep[K-S,][]{kolmogorovSullaDeterminazioneEmpirica1933,masseyjr.KolmogorovSmirnovTestGoodness1951}, the Anderson-Darling test \citep[A-D,][]{andersonTestGoodnessFit1954}, and the Cram\'er-von Mises criterion \citep{cramerCompositionElementaryErrors1928,vonmisesWahrscheinlichkeitStatistikUnd1928}.

We now take a moment to reflect on this important result. Using this statistical test, it is possible to check whether the posterior distribution is consistent with injected values over a collection of realizations of the data generation process. But this is only true if the prior distribution used in Bayes Theorem is the same as the one of the data generation process. If a different prior is used, the posterior is not going to be statistically consistent with the injections taken over \emph{many realizations of data}. If one only carries out the experiment once, on one realization of data, the posterior visually looks consistent with the injections for any reasonable prior.
But, if one analyzes many realizations, they start to be sensitive to the fact that the prior used for inference should match the distribution from which the injection parameters $\theta_{\rm{inj}}$ are drawn. It is interesting to ponder the similarity of these consistency checks to what is happening in PTAs. Because an ensemble of pulsars is analyzed simultaneously, the experiment becomes sensitive to the underlying distribution that the noise parameters of pulsars were drawn from: the collection of observable Galactic MSPs.

\begin{figure}
    \includegraphics[width=0.48\textwidth]{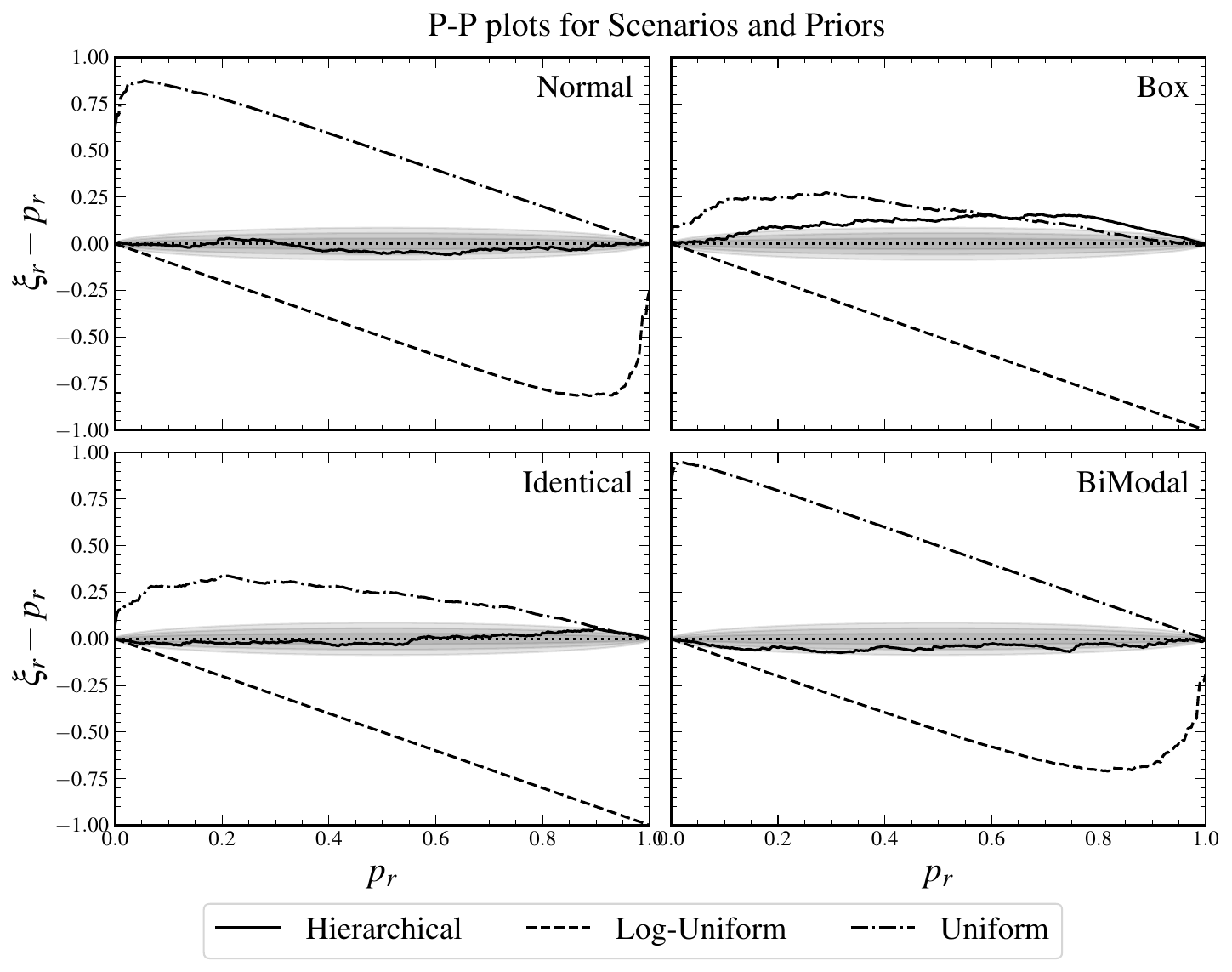}
    \caption{The four panels show the $p$-$p$ plot as calculated when generating $N_r=200$ realizations of data for the four scenarios described in Section~\ref{sec:toy_model}. The grey regions show the $1$-$\sigma$, $2$-$\sigma$, and $3$-$\sigma$ regions one would get from the Binomial distribution $\rm{Binomial(p_r, N_r)}$. The priors used for inference are the ``Hierarchical'' prior (solid), the ``Log-Uniform'' prior (dashed), and the ``Uniform'' prior (dash-dotted). The Hierarchical prior performs best overall, but it is clear from the ``Box'' scenario (which cannot be exactly modeled by a Log-Normal) that even the Hierarchical model will show slight posterior bias when the prior does not describe the data generation process. Interestingly, even the ``bi-modal'' scenario --- which is of course far from Gaussian --- seems to be \modification{well-described by the uni-model log-normal HBM.}}
    \label{fig:ppplots}
\end{figure}

Visually, $\xi_r$ and $p_r$ are often plotted against one another in a so called $p$-$p$ plot. These types of plots offer more information than just the value of a test. See for example Figure~13 of \citet{lambNeedSpeedRapid2023} for a detailed visual explanation of such plots. In this letter, we plot $\xi_r-p_r$ vs $p_r$. The mean and variance under the correct model $H$ are given by:
\begin{eqnarray}
    \langle \xi_r-p_r \rangle &=& 0  \\
    \langle (\xi_r-p_r)^2 \rangle - \langle \xi_r-p_r \rangle^2 &=& p_r(1-p_r)N_r
\end{eqnarray}
Doing this for all four scenarios of Table \ref{tab:scenarios} using the log-normal hieararchical model of Equation~\eqref{eq:hierarchical_model} gives Figure \ref{fig:ppplots}. The top-left panel shows the $p$-$p$ plots for the ``Normal'' scenario. Only the Hierarchical prior shows complete consistency with injections, which is to be expected: the generated realizations had a prior distribution that was a subset of the HBM used for analysis. Both the Log-Uniform and the Uniform priors on $\log_{10} \sigma_a$ cause significant posterior bias.
The top-right panel corresponding to the ``Box'' scenario shows that both the HBM and the Uniform priors have slight but not overwhelming posterior bias. Indeed, one can imagine that the Uniform prior is not too far off from the ``Box'' distribution that was used for the generation of data. The HBM is flexible enough to adjust to the ``Box'' prior.
The bottom-left panel shows the results for the ``Identical'' scenario. The HBM is also able to adjust to this data generation distribution, this time by reducing the variance parameter $\log_{10}\sigma_{a}$ making it approach a Dirac Delta function. Both the Log-Uniform and the Uniform show more bias. Again here, just as in the ``Box'' scenario, the Uniform priors outperform the Log-Uniform priors. Representative posteriors for all the scenarios are shown in Figure~\ref{fig:scenario-posteriors}.

\subsection{The optimal detection statistic}
\label{sec:optimalstatistic}
The optimal detection statistic \citep{anholmOptimalStrategiesGravitational2009,chamberlinTimedomainImplementationOptimal2015,polForecastingPulsarTiming2022,sardesaiGeneralizedOptimalStatistic2023a} and the pair-covariant optimal statistic \citep[][Gersbach et al. in prep.]{allenHellingsDownsCorrelation2023a} for a stochastic GWB
require an estimate or distribution of noise parameters to be reliable. Therefore, even though these Frequentist statistics are unbiased (provided no auto-correlations are used), their values do significantly depend on the noise estimates. In the PTA literature, a variant of the optimal detection statistic referred to as the noise-marginalize optimal statistic \citep[NMOS,][]{vigelandNoisemarginalizedOptimalStatistic2018} uses the samples of the posterior distribution to repeatedly evaluate the detection statistic with noise parameters from the chain. This has later been re-interpreted formally by \citet{vallisneriPosteriorPredictiveChecking2023} as posterior-prediction.
When modeling the underlying noise parameter distribution of the ensemble of observable Galactic MSPs, the noise estimates in the GWB search change, meaning that the optimal statistic values change with it. In Figure \ref{fig:scenario-posteriors}, the noise-marginalized optimal statistic distributions are shown under the different priors.

\section{Pulsar Timing examples}
\label{sec:simulations}

In a way, this article is now complete. By realizing that once an ensemble of some physical quantity is simultaneously analyzed, the underlying distribution needs to be modeled, the straightforward solution is to introduce an HBM. Of course, a suitable distribution with suitable hyper-priors on the hyperparameters should be used, such as those of Equation~\eqref{eq:hierarchical_model}. Other than that, the procedure is set.

For clarity, and to demonstrate impact, we present two more realistic examples of an HBM for pulsar timing: the First IPTA Mock Data Challenge (MDC), and an analysis of simulated data that are somewhat similar to the NANOGrav 12.5yr dataset\citep{alamNANOGrav12Yr2020}. With these examples it will become clear that there is a real need for HBMs in PTA science. But to analyze real-world data, we first need to introduce a slightly more flexible HBM
that can account for power-law signals such as those frequently modeled in PTA projects.

\subsection{Power-law PSD signals}
\label{sec:hbm2d}
In many PTA models, the Power Spectral Density (PSD) of time-correlated stochastic processes is modeled as a power-law signal:
\begin{equation}
    S(f) = A^2 \frac{c}{f_r^3} \left( \frac{f}{f_r} \right)^{-\gamma}
\end{equation}
where $A$ is amplitude of the random process, $\gamma$ is the so-called spectral index parameter, $f$ represents the signal frequency, and $f_r$ is the reference frequency often set to $f_r = \rm{yr}^{-1}$. In what follows, the log-likelihood does not change, and the assumption is made that this is calculated using standard methods in \texttt{Enterprise}. For details on how to evaluate the log-likelihood there are many detailed descriptions in the literature \citep[e.g.][]{johnsonNANOGrav15yearGravitationalWave2023a,vanhaasterenNewAdvancesGaussianprocess2014}.
In the context of the HBM, the $\log_{10}A$ should be treated as an amplitude parameter that follows some distribution, but --- different from before --- so is $\gamma$.
When using uninformative priors, the priors on these are typically set to:
\begin{eqnarray}
    \log_{10}A &\sim& \rm{Uniform}(-20, -10) \\
    \gamma &\sim& \rm{Uniform}(0, 7).
\end{eqnarray}
 Since the range of allowed $\gamma$ is bound on both sides, a log-normal distribution should be truncated. One-dimensional truncated normal distributions are easy to deal with, though perhaps a flexible distribution on the interval $[a,b]$ like the \emph{Kumaraswamy} distribution would be more appropriate if we were to model the $\gamma$ parameters separate from their $\log_{10}A$ counterparts. The Kumaraswamy distribution is defined as:
\begin{equation}
    P(x \mid a,b) = abx^{a-1}(1-x^a)^{b-1}
    \quad \text{for } 0 < x < 1,
    \label{eq:kuma}
\end{equation}
where $a,b \in \mathbb{R}^{+}$ are hyperparameters that set the shape of the distribution. However, the values of $\log_{10}A$ and $\gamma$ parameters are probably significantly correlated under the distribution of MSPs in our collection \citep{lamNANOGravNineyearDataset2016}, which means it is more realistic to use a multivariate distribution.
Instead, we therefore propose a two-step process. First, $\gamma$ is transformed $\gamma^\prime = \gamma^\prime(\gamma)$ using an interval transform, which maps the interval $[0,7] \rightarrow \langle -\infty,\infty \rangle$. Then a multivariate Gaussian prior can be placed on $(\log_{10}A,\gamma^\prime)$.

The interval transform is defined as:
\begin{eqnarray}
    \gamma^\prime &=& \log \left( \frac{\gamma - a}{b - \gamma} \right) \\
    \gamma &=& \frac{(b-a)\exp (\gamma^{\prime})}{1+\exp (\gamma^{\prime}})\\
    \frac{\rm{d}\gamma^{\prime}}{\rm{d}\gamma} &=& \frac{b-a}{(b-\gamma)(\gamma-a)}
    \label{eq:intervaltransform}
\end{eqnarray}
The Jacobian $\rm{d}\gamma^\prime / \rm{d} \gamma$ is needed to transform probability densities. In \texttt{PyMC} transformations are done automatically behind the scenes, but it is quite straightforward to include all necessary terms manually. Note, though, that coordinate transformations like this can seriously distort point estimators like the HBM Maximum A Posteriori (MAP). When constructing a point estimator of an HBM, it makes sense to \emph{not} include the Jacobian.

We now define Multivariate Gaussian prior $\mathcal{N}(\mu_{h},\mathbf{\Sigma_{h}})$ ($h$ for 'hyper') as a function of $5$ hyperparameters: two mean parameters $\mu_{h}=(\mu_{\log_{10}A}, \mu_{\gamma^{\prime}})$, and three variance parameters $L_{A}, L_{\gamma}, L_{A,\gamma}$. The two mean parameters $\mu_{h}$ are the mean of the Multivariate Gaussian. The three $L$ parameters parameterize the non-zero elements of the Cholesky decomposition of the covariance matrix $\mathbf{\Sigma}_{h}$:
\begin{eqnarray}
    \mathbf{L} &=& \begin{bmatrix}
        L_{A} & 0 \\ \nonumber
        L_{A\gamma} & L_{\gamma} \nonumber
    \end{bmatrix} \\
    \mathbf{\Sigma}_{h} &=& \mathbf{LL}^T
\end{eqnarray}
Where $\mathbf{\Sigma}$ is the covariance matrix of the multivariate Gaussian.
We place the following priors on these hyperparameters:
\begin{eqnarray}
    \mu_{\log_{10}A} &\sim& \rm{Uniform}(-20, -13) \nonumber \\
    \mu_{\gamma^\prime} &\sim& \rm{Uniform}(-10, 10) \nonumber \\
    L_{A} &\sim& \rm{Uniform}(0.3, 2.3) \\
    L_{A\gamma} &\sim& \rm{Uniform}(-1.5, 1.5) \nonumber \\
    L_{\gamma} &\sim& \rm{Uniform}(0.3, 2.3) \nonumber
    \label{eq:hyperpriors}
\end{eqnarray}
These ranges are chosen somewhat arbitrarily, though they are sufficient for the datasets that are analyzed here. By not allowing $L_{A}$ and $L_{\gamma}$ to go lower than $0.3$, some low mixing rate of the MCMC chains are avoided.
Regardless, these hyperparameters parameterize the distribution of the Galactic population of observable MSPs, and there is considerable variability among pulsar spin noise, meaning the diagonal $L$ components should not go towards $0$. The spin noise parameters $x^{\prime} = (\log_{10}A, \gamma^\prime)$ now have the Multivariate Gaussian prior:
\begin{equation}
    x^{\prime} \sim \mathcal{N}(\mu_{h}, \mathbf{\Sigma}_{h}).
    \label{eq:hyperprior2d}
\end{equation}

A note that is in order is that it can be challenging to sample from Hierarchical models like these due to the extreme covariance of low-level parameter with high-level parameters. This is especially true for variance parameters, like $L_{A}$, which greatly restrict parameter ranges of low-level parameters. As discussed in the influential paper on slice sampling by \citet{nealSliceSampling2003}, such parameter funnels are best avoided using coordinate transformations. The proper decentering transformation that should be used for this HBM is
\begin{equation}
    \label{eq:decentering}
    x^{\prime\prime} = \mathbf{L}^{-1}(x^{\prime} - \mu_{h}).
\end{equation}
Using these coordinates as the sampling parameters greatly increases the MCMC mixing rate, and allows traversing otherwise inaccessible regions of parameter space. As we noted before, if using transformations such as these, remember to include the Jacobian. The final coordinate transformation chain then becomes:
\begin{equation}
    x \rightarrow x^{\prime} \rightarrow x^{\prime\prime}.
\end{equation}
The low-level parameters as they are used in \texttt{Enterprise} are denoted with $x$. Those parameters are then transformed (only the spectral indeces of the noise) to $x^{\prime}$ using Equation~\eqref{eq:intervaltransform}. After that, the low-level parameters are transformed to $x^{\prime\prime}$ using Equation~\eqref{eq:decentering}.

It is more than likely that Hierarchical models exist that better fit PTA data than the relatively straightforward model introduced in this section. Just like what is already done with advanced noise models, various HMBs will need to be explored to see what works best by the community. For instance, the coordinate transformation $\gamma \rightarrow \gamma^\prime$ has a divergence of the Jacobian near the edges that can be challenging for certain numerical methods. And perhaps a multivariate normal is not appropriate for these parameters, and instead multi-modal and asymmetric distributions can serve as extensions to the population model introduced above. In fact, a power-law spectral density is likely an incomplete description for time-correlated noise in MSPs in general. The entire HBM will need to be reconsidered as the community starts to study the population of MSPs that are regularly observed as part of timing programs across the globe.

\subsection{First IPTA Mock Data Challenge}
\label{sec:mdc1}

The first IPTA MDC (MDC1) included three open and three closed sets of data, where for the open sets the injection sources and parameters were known, and for the closed sets no information was released. The open sets only included white (IID) noise and an injection of a GWB with a power-law PSD and known spectral index. No spin noise was added to the data of the three open sets. Multiple research groups submitted an analysis of the three closed datasets \citep{vanhaasterenAnalysisFirstIPTA2013,cornishApplyingBayesianInference2012,ellisResultsFirstIPTA2012a,taylorWeighingEvidenceGravitationalWave2013, lentatiHyperefficientModelindependentBayesian2013}, some of which assumed that the closed sets might include added spin noise, even though the open sets did not. The used priors by these groups can be summarized as
\begin{align}
    \log_{10}h &\sim \rm{Uniform}\left(-20, -10\right) \nonumber \\
    \log_{10}n_{a} &\sim \rm{Uniform}\left(-20, -10\right)    & \text{log-uniform} \\
    n_{a} &\sim \rm{Uniform}\left(0, 10^{-10}\right)          & \text{uniform} \nonumber
\end{align}
where the last two lines give the two different options for spin noise amplitude: either log-uniform priors, or uniform priors. The parameters $log_{10}n_{a}$ represent the spin noise amplitude of pulsar $a$ in the same units that \texttt{Enterprise} uses. In those units, the spin noise is basically defined as a pulsar-term GW signal for only pulsar $a$. For details on how the likelihood and posterior are formed for realistic data, we refer to \citet[][and references therein]{vanhaasterenNewAdvancesGaussianprocess2014}.

\begin{figure}
    \includegraphics[width=0.48\textwidth]{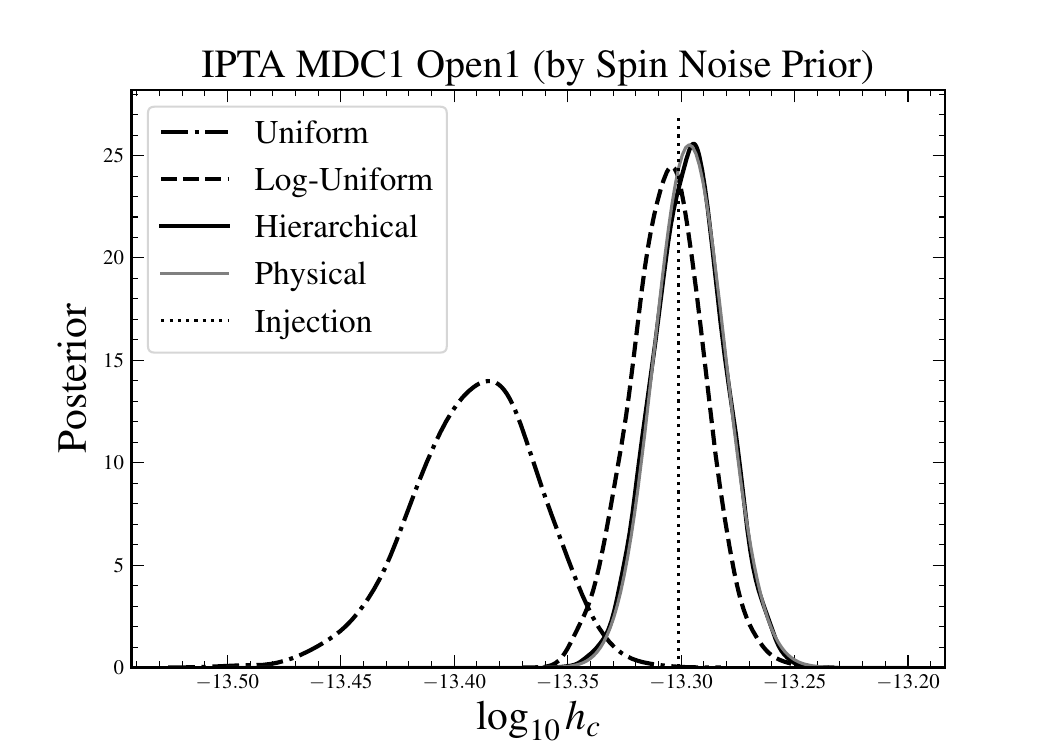}
    \caption{The posterior distribution of the GWB characteristic strain $h_c(\rm{yr}^{-1})$ of the First IPTA MDC, open dataset 1. Free model parameters were all timing model parameters, spin noise amplitude for all pulsars, and the GWB amplitude $h_c$. Different lines are for different priors on the spin noise amplitudes that have been marginalized over. The ``Physical'' prior means the spin noise amplitudes have been set to their injected value: 0. Visually the hierarchical prior approximates the posterior of the Physical model the most. The posterior for Uniform priors is visually inconsistent with the injection parameters.}
    \label{fig:mdc1-open1-1d}
\end{figure}

Figure \ref{fig:mdc1-open1-1d} shows the results of a search for $h$ and $n_{a}$ in the MDC1 open1 dataset. Clearly, the results are inconsistent with one another, and the prior that is uniform in $n_{a}$ is inconsistent with the injections.

From the MDC1 onward, the log-uniform priors were widely accepted as the ``correct'' choice for spin noise and many other noise processes for detection, with the acknowledgement that these priors are effectively improper and need arbitrary bounds. When upper-limits on a parameter need to be set, current software packages allow somewhat arbitrarily to switch to uniform priors so that the lower bound of amplitude parameters is well-defined.
In Figure \ref{fig:mdc1-open1-1d} we have also plotted the posterior distribution that is obtained with a model \emph{without} any spin noise, labeled as ``Physical''. The full posterior here is only one-dimensional, meaning that no marginalization or sampling was required to make it. It is clear that, although the ``Log-Uniform'' and ``Physical'' priors give comparable results, the posterior made with the ``Physical'' prior is shifted slightly to the right.

The hierarchical model we used for the MDC was:
\begin{eqnarray}
    \log_{10}\mu_{n} &\sim& \mathcal{N}(-19, 5^2) \nonumber \\
    \log_{10}\sigma_{n} &\sim& \rm{Uniform}(0.1, 3.1) \\
    \log_{10}n_{a} &\sim& \mathcal{N}\left(\log_{10}\mu_{n}, (\log_{10}\sigma_{n})^2\right), \nonumber
    \label{eq:hierarchical_model_mdc}
\end{eqnarray}
where the hyperparameters $\log_{10}\mu_{n}$ and $\log_{10}\sigma_n$ are given some broad priors that do not make the sampling too inefficient. The posterior that results from this prior is shown in Figure \ref{fig:mdc1-open1-1d}, the line of which lies on top of/underneath the posterior for the ``Physical'' prior. The two are indistinguishable, which is reassuring that the Hierarchical prior is doing what it is supposed to be doing.

This example demonstrates that both uninformative priors that were applied to the
IPTA MDC1 datasets in the literature were sub-optimal in hindsight. It just so
happened that the Log-Uniform priors were not wrong enough for anyone to notice
until now. Using nested sampling with \texttt{Dynesty}
\citep{speagleDynestyDynamicNested2020}, we have determined that the Bayes factor
$\mathcal{B} = Z_{\rm{Hierarchical}} / Z_{\rm{Log-Uniform}} \approx 1665$ is
significantly in favor of the Hierarchical model.
Note that, since MDC1 did not
contain any injected spin noise, the likelihood is effectively ``high'' in a
large portion of prior space: everywhere under the $\log_{10}n_a \le -13.7$ the
likelihood value is somewhat similar, since that is small with respect to the
injected GWB. If spin noise were present and detectable, like it is in realistic
datasets, the Bayes factors between models with Hierarchical and Log-Uniform
priors can get drastically larger.

\subsection{More realistic mock data}
\label{sec:ng15_scenario}

\begin{figure}
    \includegraphics[width=0.48\textwidth]{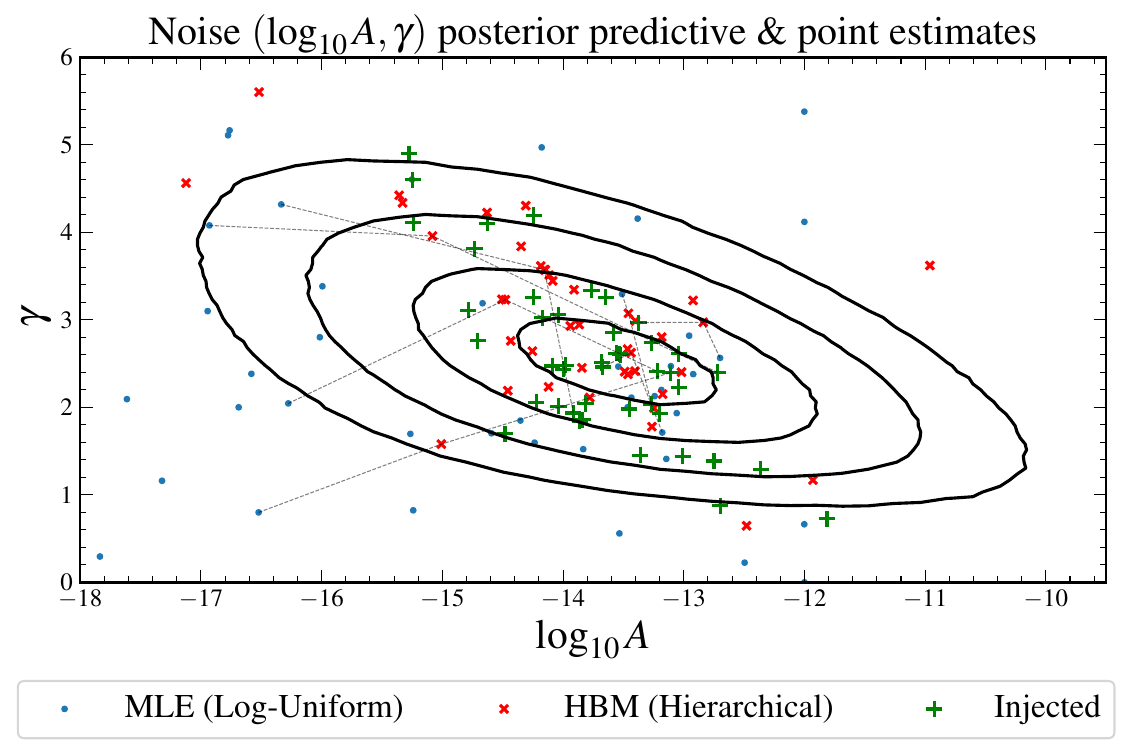}
    \caption{Point estimators for the spin noise parameters $(\log_{10}A, \gamma)$ for the Log-Uniform (Maximum Likelihood Estimator, MLE) and the HBM model, plotted together with the injected values. The Posterior Predictive for $(\log_{10}A, \gamma)$ is shown as contours. Although the point estimates were not very stable numerically due to light degeneracies in the model, the effects of shrinkage can be observed as the HBM point estimates are drawn towards the middle of the posterior predictive. For some pulsars we have also drawn lines between the MLE, HBM, and injection values, so that their relationship can be seen.}
\label{fig:postpred}
\end{figure}

As a last more realistic test, we generated some mock data that was loosely modeled after the NANOGrav 12.5yr dataset \citep{alamNANOGrav12Yr2020}, which showed a significant detection of a common process in the GWB search \citep{arzoumanianNANOGrav12Yr2020}. Correlations were not significantly observed, but a model with a common uncorrelated process (CURN) was preferred over noise-only with a Bayes factor $\mathcal{B} = Z_{\rm{CURN}} / Z_{\rm{Noise}} > 10^4$.  We created a custom \texttt{Enterprise} class for mock data so that no actual PTA datasets were required as input. As timing model we used quadratic spindown and astrometry parameters, the latter of which were created using \texttt{Astropy} \citep{astropycollaborationAstropyProjectSustaining2022}\footnote{Code to generate mock data as an Enterprise addition: \\ \url{https://github.com/nanograv/enterprise/pull/361}}. Time of observations and spin noise parameters were loosely based on the NANOGrav 12.5yr datasets.
Spin noise parameters $\log_{10}A$ and $\gamma$ are typically covariant due to the choice of reference frequency $f_r$. We generated the spin noise parameters loosely scattered around the line of covariance with total timing residual variance based on the values in the literature. The exact values are shown in Figure~\ref{fig:postpred}.
For the GWB, we injected a $\log_{10}h_{c}=-15$ signal that is correlated among pulsars with the regular Hellings \& Downs correlations $\Gamma(\zeta)$.

\begin{figure}
    \includegraphics[width=0.48\textwidth]{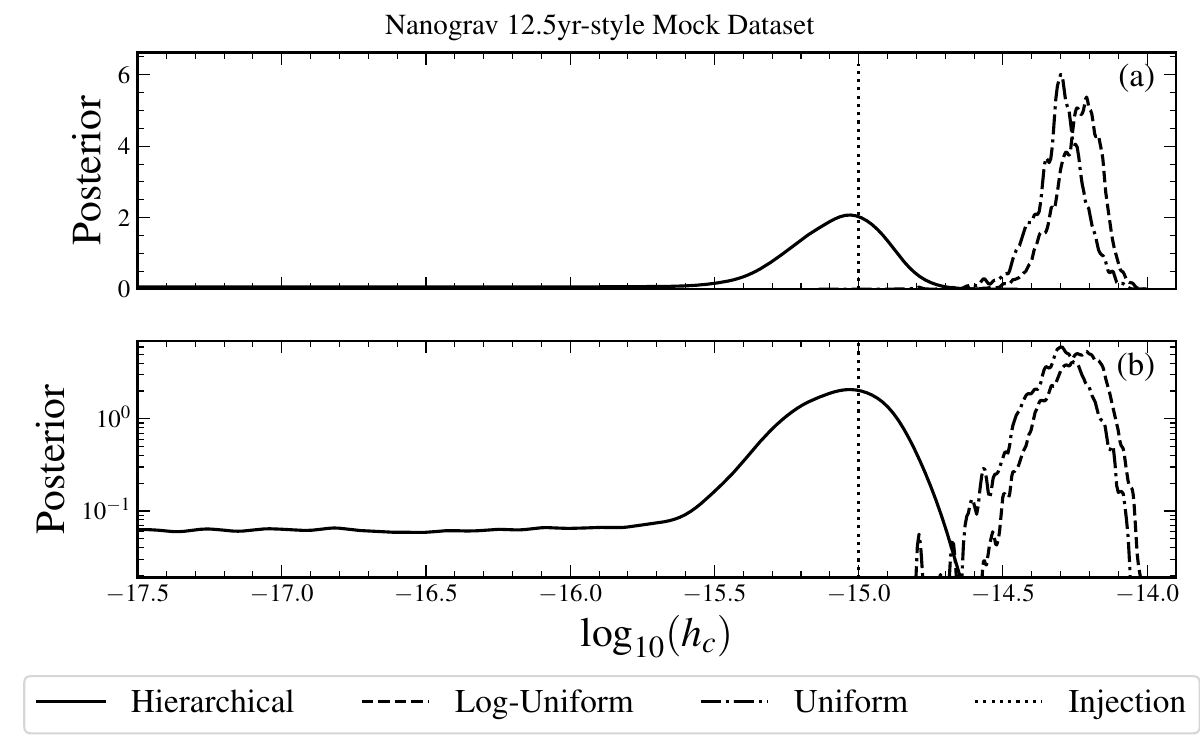}
    \caption{The posterior distribution of the GWB characteristic strain $h_c(\rm{yr}^{-1})$ of a mock dataset loosely modeled after the NANOGrav $12.5$yr dataset. The injected spin noise came from a distribution of plausible parameters, and the injected GWB amplitude was conservatively set to $h_c=10^{-15}$. The top and bottom plot show the same graph, but the scaling of the y-axis is linear and logarithmic for respectively the top and bottom panel. The model with Hierarhical priors (solid) for spin noise shows consistency with the injection, whereas the model with Log-Uniform (dashed) and Uniform priors (dash-dotted) for spin noise display significant posterior bias. The amount of posterior bias on display here is reason for concern regarding current PTA analyses in the literature.}
    \label{fig:ng12mock-posterior}
\end{figure}

Using a different random number seed at the start of the data generation, we created several realizations to see how often we would get a discrepancy between a model with Log-Uniform priors and Hierarchical priors. It turned out that occasionally both models were in agreement, and sometimes they did not. The toy model analysis from Section~\ref{sec:toy_model}, which was built completely independently with \texttt{PyMC}, exhibited the same behavior. For the mock data in this Section, we noticed a serious discrepancy between the model with Log-Uniform priors and the model with Hierarchical priors on the \emph{third} realization of mock data, which gives an impression of how often such posterior bias occurs.
The analysis result of that mock dataset is shown in Figure~\ref{fig:ng12mock-posterior}. It is clear that only the model with Hierarchical priors managed to recover the injection here. Statistical analysis coupled with more proper noise modeling of realistic PTA data is postponed to future investigations. The goal of this example is to strengthen the statement made in the title of this letter: posterior bias is not rare, and an HBM can address it.

To show the effect of the HBM on the recovery of spin noise parameters we constructed point estimators of $\log_{10} A$ and $\gamma$ for each pulsar under the different priors, as can be seen in Figure~\ref{fig:postpred}. For the Log-Uniform priors, we used the Maximum Likelihood Estimator (MLE, marked with 'o') on the likelihood from \texttt{Enterprise}. For the HBM (marked with 'x'), we used the log-posterior distribution minus the log-Jacobian to avoid artifacts coming from the coordinate transformation around the edges of the allowed interval for the $\gamma$ parameters. We also plotted the injected values (marked with '+').
The effect of shrinkage can be seen, where the HBM estimates in general are more clustered towards the middle of the distribution. One caveat with these results is that the CURN model is lightly degenerate with respect to the model parameters, so the optimization results are not very stable numerically. Some point estimates have a significant margin of error. Regardless, the shrinkage effect can be observed.

\section{Comparison with attempts in the literature}
\label{sec:litcomparison}
Several studies in the literature have attempted to explain potential biases in PTA results. Firstly, \citet{zicEvaluatingPrevalenceSpurious2022} pointed out that spurious detections could happen in simulated datasets that did not have a common signal injected. Similarly, \citet{johnsonGravitationalwaveStatisticsPulsar2022} point out that biases can happen when analyzing PTA data, and they evaluate many realizations of data to understand when this is happening.
\citet{goncharovConsistencyParkesPulsar2022} remedy the model misspecification found by \citet{zicEvaluatingPrevalenceSpurious2022} by including a ``quasi-common process'', which is similar to the common process (GWB) but with amplitudes governed by hyperparameters. Uninformative priors were left in place for noise. Such a model seemed to resolve the posterior bias they found, but it is still misspecified from a population point of view.
Lastly, \citet{hazbounModelDependenceBayesian2020} found posterior bias in their simulations, and they discovered that a ``dropout'' model, that allows the data to decide whether the model should include a spin noise component for a pulsar, can reduce the posterior bias in the GW parameters. The approach of turning off the spin noise of a pulsar (as if removing it from the ensemble) \modification{based on the single-pulsar analysis odds ratio is} incorrect. \modification{This is because this process is equivalent to using the data to change the  prior odds of various noise models, and then re-analyzing said data with updated prior odds informed by the data. In the Bayesian Statistics literature having a data-informed prior is referred to as \emph{double-dipping}, and it is not allowed.} \modification{In general},

\begin{principle}
\modification{The most correct and conservative model contains all signal contributions that could have a measurable effect under the prior.}
\end{principle}

\modification{That being said}, letting the data decide whether a pulsar \modification{needs certain noise components modeled} is not unreasonable, especially if \modification{those noise processes are not covariant with the signals of interest}. \modification{It is possible to explain the dropout model in that context, but it depends on the exact model and data details, which is outside the scope of this paper.}

All the above treatments in the literature analyzed their simulated datasets with a different prior distribution than what was used for data generation. This alone explains the posterior bias these analyses exposed. Of course, their assessment was correct in that this posterior bias was an important issue that needed to be addressed.

\section{Implications for Pulsar Timing Array projects}
\label{sec:implications}
As outlined in Section~\ref{sec:ppplots}, the Hierarchical priors should make for more robust results on GW parameters with less posterior bias in general, even when we do not know well what the underlying distribution of the noise parameters of the ensemble of observable Galactic MSPs is. Conversely, GW results obtained with models that have Log-Uniform or other uninformative priors may need to be re-evaluated since all results from PTA projects in 2023 \citep{agazieNANOGrav15Yr2023b,reardonSearchIsotropicGravitationalwave2023,antoniadisSecondDataRelease2023e, xuSearchingNanoHertzStochastic2023} have used such priors.

The observed correlations reported by the various PTA projects, such as those calculated with the Optimal Statistic \citep{anholmOptimalStrategiesGravitational2009,chamberlinTimedomainImplementationOptimal2015, allenHellingsDownsCorrelation2023a, vigelandNoisemarginalizedOptimalStatistic2018, sardesaiGeneralizedOptimalStatistic2023a,vallisneriPosteriorPredictiveChecking2023}, also depend on the noise parameters. However, even with biased noise parameter values, those statistics are still unbiased.
Further, the false alarm probabilities created using simulations based on the null-Hypothesis should still hold if the background estimation method from sky scrambles and phase shifts \citep{cornishRobustGravitationalWave2016,taylorAllCorrelationsMust2017} are valid. Although this has not been formally derived \citep{dimarcoRobustDetectionsNanohertz2023}, simulations show that such background estimates are quite reasonable (Taylor, private communications).
The concerned reader can therefore have some peace of mind that the evidence for a GWB in PTA data is not going to go away.
Regardless, even the Noise Marginalized Optimal Statistic will have to be recalculated with updated noise parameters. It is unclear what results will change to when appropriate HBMs are crafted for real PTA data. This will be a community effort, and the model explored in this letter is almost surely inadequate \modification{in some way. Regardless, because the model has been misspecified when uninformative priors are used on parameters that are covariant with the GWB, the inference on GWB parameters will change under an HBM}. In summary, one should prepare for updated Bayes Factors, modified estimates of the characteristic strain $h_c$ for power-law models, updated GWB spectral estimates in general, and potentially slightly altered false-alarm probabilities.

\section{Conclusions}
\label{sec:conclusions}
GW searches by PTAs use an array of pulsars which are jointly analyzed. Many noise sources are modeled simultaneously with the GW signal. The noise source processes originating from individual pulsars are separately characterized in such an analysis, and the parameters of these processes are typically given an uninformative prior of some sort. Such uninformative priors should not be used when an ensemble of processes is modeled jointly, and instead the underlying distribution of noise parameters of the ensemble of pulsars should be modeled simultaneously with all the pulsar properties.
This joint model can be elegantly described by a Hierarchical Bayesian Model, which is the combination of the original likelihood with a different Hierarchical Prior. Neglecting to model the underlying distribution in such a way biases model parameters. Especially GW parameters are affected, since the GWB is somewhat covariant with the whole ensemble of spin noise parameters. Current results in the PTA literature all use uninformative priors for the pulsar spin noise, and should be re-evaluated in the context of hierarchical models. This could result in updated Bayes Factors, different (GWB) spectral estimates, and slightly modified false alarm probabilities.

Not only pulsar spin noise parameters can be modeled with a Hierarchical Model, but any process and parameter of the ensemble of pulsars that is currently modeled in the PTA can (and perhaps \emph{should}) be modeled with a Hierarchical Model. For example, the Gaussian Process Dispersion Measure Variations model is an obvious candidate for a Hierarchical approach, and also certain timing model parameters could conceivably benefit from a Hierarchical Model.

\begin{acknowledgments}
I thank Pat Myers and Boris Goncharov for useful discussions regarding priors in Bayesian analysis.
\end{acknowledgments}

\appendix

\section{Extra discussion of the toy model bi-modal scenario}
\begin{figure}[h]
    \includegraphics[width=0.95\textwidth]{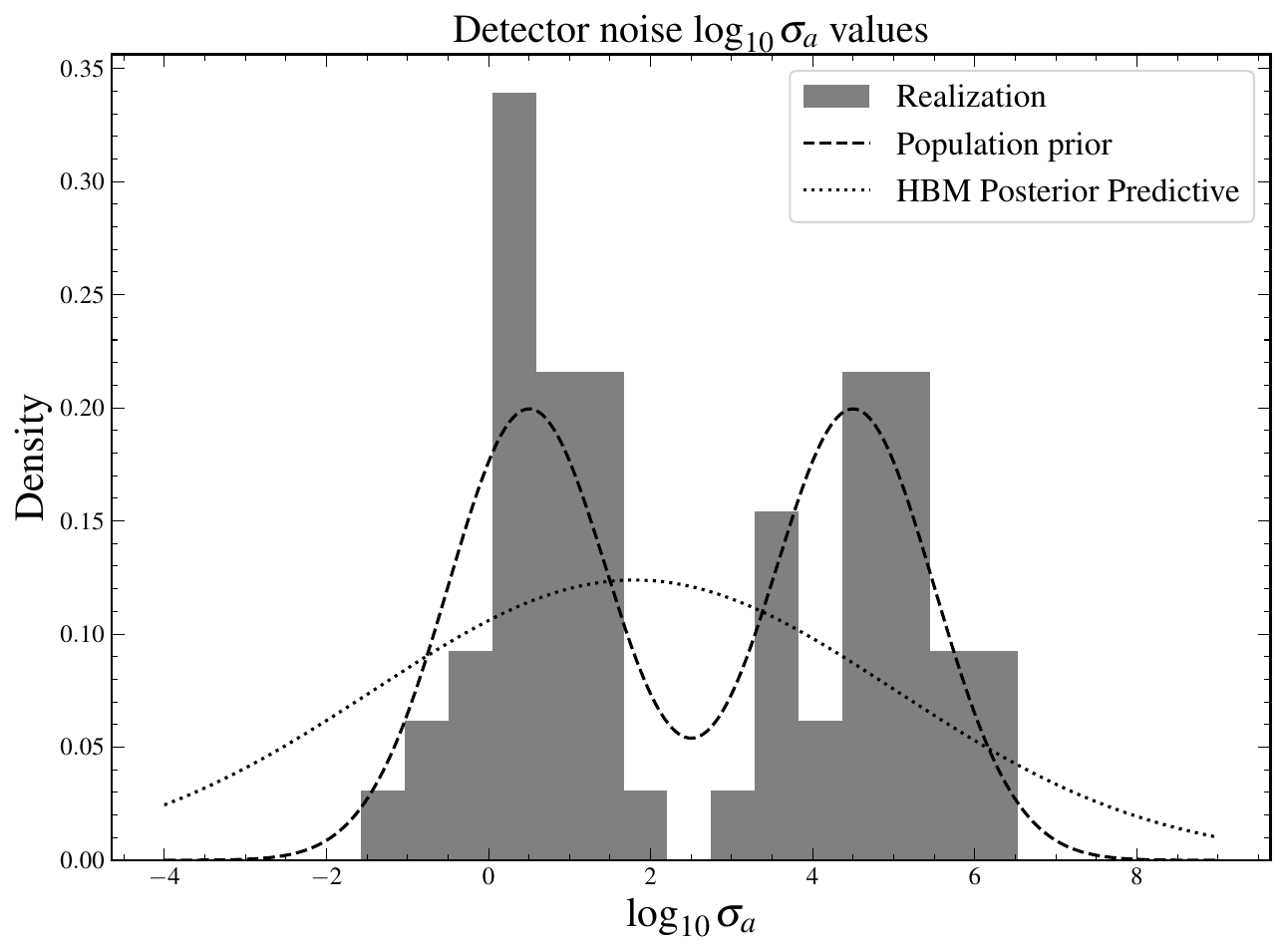}
    \caption{The detector noise values in the realization of data represented by the bi-model scenario of Figure~\ref{fig:scenario-posteriors}. The solid grey histogram shows the true noise values of the population. The dashed line represents the true underlying population prior, from which the values in the histogram were drawn in this realization of the universe. The dotted line shows the (median) posterior predictive as inferred by the HBM: it is a Gaussian with both the mean and variance picked from the MCMC chain at their 50th percentile.}
    \label{fig:bimodal-noise-dist}
\end{figure}

The bi-modal distribution that was used as part of the four toy model scenarios is insightful to look at in more detail: the underlying population prior is highly non-Gaussian, and one of the two modes of the population represents detectors with largely undetectable noise. Figure~\ref{fig:bimodal-noise-dist} shows the underlying population prior that is also listed in Table~\ref{tab:scenarios} for the bi-modal scenario as a dashed line. The histogram shows the detector noise values in the realization of the scenario that is also shown in Figure~\ref{fig:scenario-posteriors}. The dotted line shows the posterior predictive under the HBM, taken at the 50th percentile of the hyperparameters $\mu_\sigma$ and $\sigma_\sigma$.
It is clear from Figure~\ref{fig:bimodal-noise-dist} that the HBM does not capture the underlying distribution well, as the two distributions are visually very different. However, as shown in Figure~\ref{fig:ppplots}, this does not cause large posterior bias on average when taken over many realizations. Intuitively, the HBM restricts the parameter space where necessary, but the prior is still ``soft'' enough that the resulting posterior distribution for the noise parameters is not pulled far away from the value that is informed by the individual detector data.

\begin{figure}[h]
    \begin{minipage}[b]{0.5\linewidth}
      \centering
      \includegraphics[width=\linewidth]{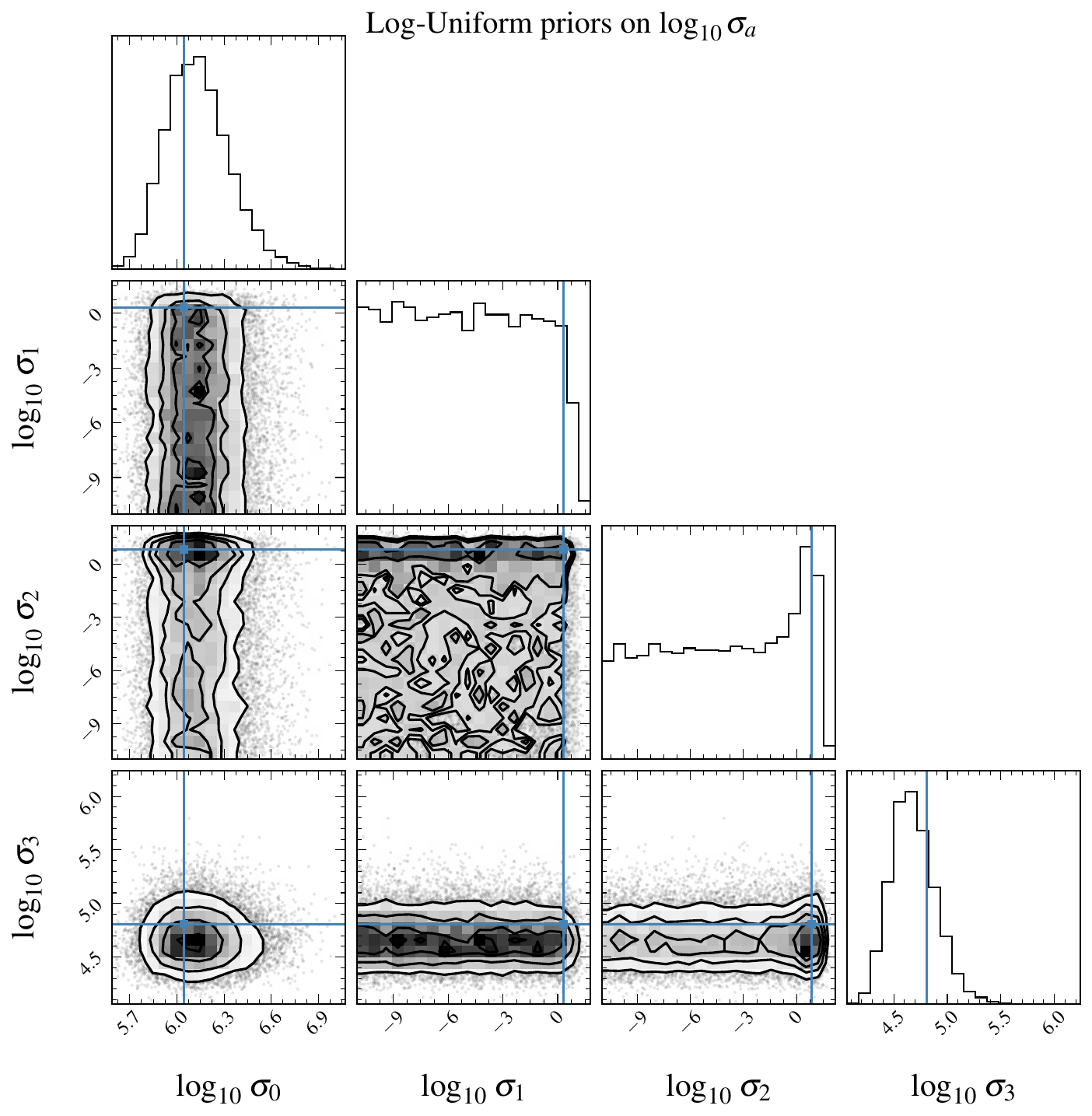}
      \caption{The detector noise $\sigma_a$ posterior distributions for the first four detectors when using Log-Uniform priors. True injected values are shown by the lines/cross. Detector noise $\sigma_0$ and $\sigma_3$ are detectable, whereas $\sigma_1$ and $\sigma_2$ are not. Note the difference in scale of the axes. The posterior distributions for $\sigma_1$ and $\sigma_2$ show support all the way to the lower boundary of the prior, since the data are consistent with no detector noise.}
      \label{fig:bimodal-noise-corner-loguni}
    \end{minipage}
    \hspace{0.5cm}
    \begin{minipage}[b]{0.5\linewidth}
      \centering
      \includegraphics[width=\linewidth]{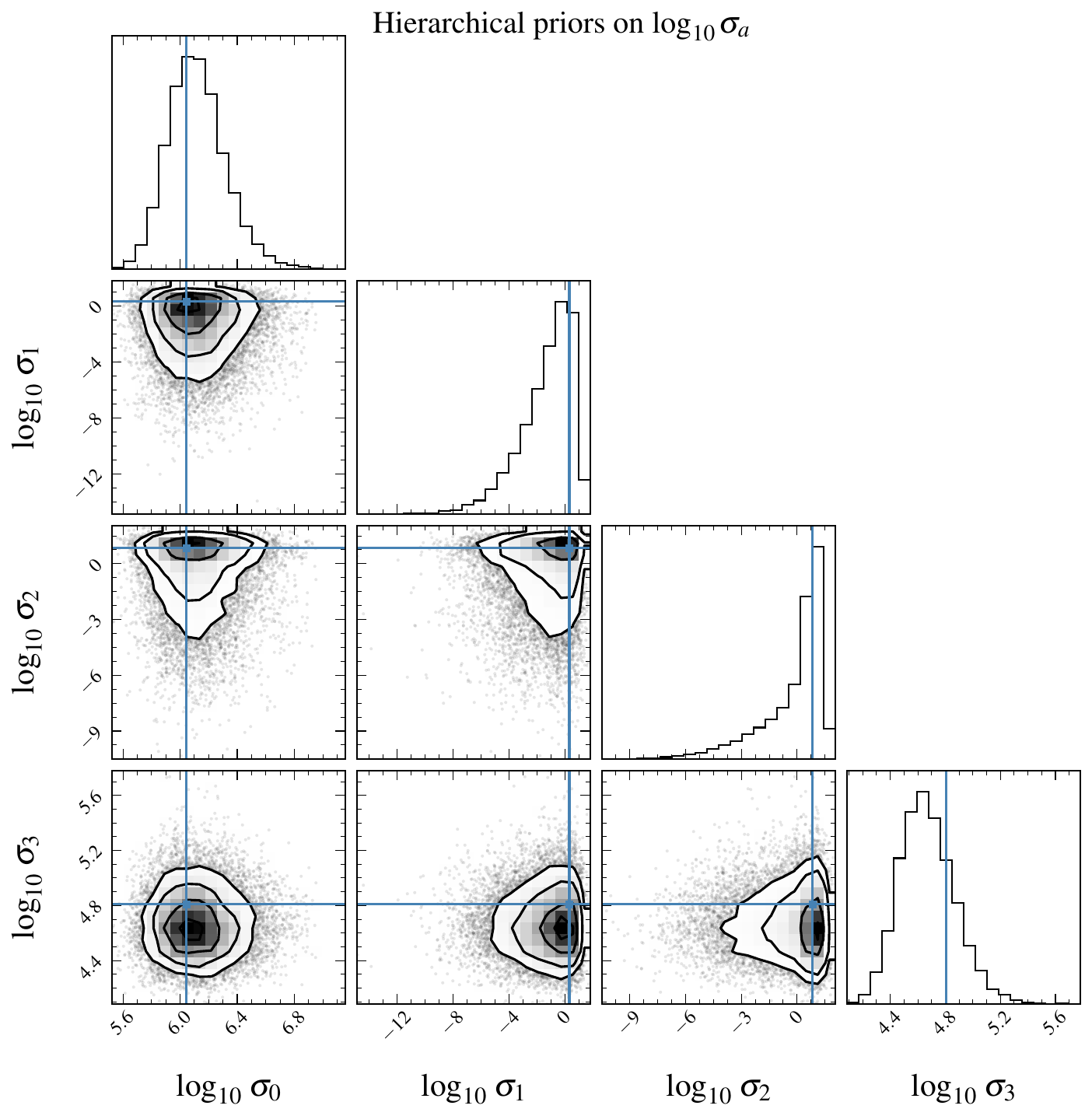}
      \caption{The detector noise $\sigma_a$ posterior distributions for the first four detectors when using hierarchical priors. True injected values are shown by the lines/cross. Compared to Log-Uniform priors, we now see that the estimates for $\sigma_1$ and $\sigma_2$ have borrowed strength from the ensemble, whereas the posterior distributions for $\sigma_0$ and $\sigma_3$ remain largely unchanged from the Log-Uniform case.
      }
      \label{fig:bimodal-noise-corner-hbm}
    \end{minipage}
  \end{figure}

The above can be visualized by looking at the noise parameters $\sigma_a$ for the Log-Uniform model and the HBM for the same realization of data. This is done in Figure~\ref{fig:bimodal-noise-corner-loguni} \& \ref{fig:bimodal-noise-corner-hbm}. What is shown here is the corner plot for the detector noise parameters $\sigma_a$ for the first four detectors. The parameters $\sigma_0$ and $\sigma_3$ come from the large mode on display in Figure~\ref{fig:bimodal-noise-dist}, whereas $\sigma_1$ and $\sigma_2$ come from the small mode. The corner plots show that under the Log-Uniform prior the posterior for $\sigma_1$ and $\sigma_2$ is unconstrained on the low end: the posterior has support all the way to the low-value edge of the prior.
Under the HBM, the posterior for $\sigma_1$ and $\sigma_2$ is constrained on either end, but it is also visible that the HBM does not push the values of $\sigma_1$ and $\sigma_2$ to regions that are not supported by the data compared to the Log-Uniform prior. It is also visible that the posterior distributions for $\sigma_0$ and $\sigma_3$ are roughly identical between the Log-Uniform prior and the HBM.

All this is reassuring. The HBM does restrict the parameter space for the individual detector paramters $\sigma_a$, and the estimates do indeed borrow strength from the ensemble, as evidenced by the posteriors of $\sigma_1$ and $\sigma_2$ under the HBM. However, the p-p plots of Figure~\ref{fig:ppplots} and the posterior distributions of $\sigma_a$ also show that the estimates are still consistent with the data.

\bibliographystyle{aasjournal}
\bibliography{references}

\end{document}